\documentstyle[psfig]{l-aa}
	
\begin{document}

\thesaurus{03(04.19.1;11.01.2;11.17.2;11.19.1;13.25.2)}

\title{The ROSAT Deep Survey}

\subtitle{II. Optical identification, photometry and spectra of 
X-ray sources in the Lockman field}

\author{M. Schmidt \inst{1} \and G. Hasinger \inst{2} \and
        J. Gunn \inst{3} \and D. Schneider \inst{4} \and
        R. Burg \inst{5} \and R. Giacconi \inst{6} \and
        I. Lehmann \inst{2} \and J. MacKenty \inst{7} \and
        J. Tr\"umper \inst{8} \and G. Zamorani \inst{9,10}  }

\offprints{G. Hasinger}

\institute{California Institute of Technology, Pasadena, CA 91125, USA
\and Astrophysikalisches Institut Potsdam, An der Sternwarte 16,
     14482 Potsdam, Germany
\and Princeton University Observatory, Princeton, NJ 08540, USA
\and Pennsylvania State University, University Park, PA 16802, USA
\and Johns Hopkins University, Baltimore, MD 21218, USA
\and European Southern Observatory, Karl-Schwarzschild-Str. 1,
     85748 Garching bei M\"unchen, Germany
\and Space Telescope Science Institute, 3700 San Martin Drive,
     Baltimore, MD 21218, USA
\and Max-Planck-Institut f\"ur extraterrestrische Physik,
     Karl-Schwarzschild-Str. 2, 85740 Garching bei M\"unchen, Germany
\and Osservatorio Astronomico, Via Zamboni 33, 40126 Bologna, Italy
\and Istituto di Radioastronomia del CNR, via Gobetti 101,
     I-40129, Bologna, Italy}
  
\date{Received 12 May 1997; accepted 11 Sep 1997}

\maketitle

\begin{abstract}

The ROSAT Deep Survey includes a complete sample of 50 X-ray sources 
with fluxes in the $0.5 - 2$ keV band larger than 
5.5\,$10^{-15}$ erg cm$^{-2}$ s$^{-1}$ in the {\em Lockman} field 
(Hasinger et al., Paper I). We have obtained deep broad-band
CCD images of the field and spectra of many optical objects near
the positions of the X-ray sources. We define systematically the
process leading to the optical identifications of the X-ray sources. 
For this purpose, we introduce five identification (ID) classes that 
characterize the process in each case.  
Among the 50 X-ray sources, we identify 39 AGNs, 3 groups of 
galaxies, 1 galaxy and 3 galactic stars. Four X-ray sources remain
unidentified so far; two of these objects may have an unusually large
ratio of X-ray to optical flux.


\end{abstract}

\section{Introduction} 

The goal of the ROSAT Deep Survey (RDS) is to obtain information 
about the luminosity functions of various types of X-ray
sources as well as their evolution with redshift, and to understand
the origin of the X-ray background (XRB) discovered more than thirty
years ago (Giacconi et al. 1962). The survey was conducted 
in the {\em Lockman} field (Hasinger et al. 1993), which has a minimum 
of galactic hydrogen column density (Lockman et al. 1986).
The survey consists of exposures totaling 207 ksec with the ROSAT PSPC 
and a raster scan of HRI exposures totaling 205 ksec. In addition,
we have a series of HRI exposures of a fixed area within the
PSPC field with a total exposure of 1112 ksec, which  allows us
to derive improved positions for PSPC sources detected in the area.  

The extraction of the X-ray sources and the derivation
of the limiting X-ray flux of detection and other X-ray properties
are presented in Paper I (Hasinger et al. 1997). 
Radio observations of the {\em Lockman} field have been published by
De Ruiter et al. (1997).

This paper presents the optical observations made in the {\em Lockman}
field. These consist of CCD imaging of the field and spectroscopic
observations of optical objects close to the position of the X-ray
sources. Section 2 covers the CCD imaging, the extraction
of the optical objects, their positions and magnitudes. 
In Sect. 3, we discuss the
spectroscopic observations of optical candidates for identification
with the X-ray sources. On the basis of these observations
and the X-ray properties, we obtain optical identifications and 
classifications for almost all of the X-ray sources, cf. Sect. 4.
In Sect. 5, we compare our results to those of other
surveys made with ROSAT and in Sect. 6 we summarize the results.
Throughout the paper, we use
$H_o =$ 50 km s$^{-1}$ Mpc$^{-1}$ and $q_o =$ 0.5.

\section{Optical imaging and photometry of the Lockman field} 

As described below, we carried out imaging observations of the
field at Mauna Kea and at Palomar.
Efforts to observe the field at Kitt Peak National Observatory
were unsuccessful due to weather.

\subsection{CCD imaging at the University of Hawaii 2.2-m telescope} 

Images of the {\em Lockman} field were obtained on 16-19 January,
1988 with the University of Hawaii 2.2-m telescope. The detector
was an 800x800 thinned Texas Instruments CCD obtained from the NSF
distribution. A focal reducer in the cassegrain camera produced an
image scale of $0\farcs355$ per pixel; the field of view was
$4\farcm73$ on a side.
A mosaic of 50 frames was taken through B and R filters on the 
Kron-Cousins system to cover 
most of the area in the inner $20\arcmin$ of the PSPC field.
Exposure times were 10 min. for the $B$ filter, and 4 min. for $R$.
Photometric calibration was provided each night by several 
observations of the M67
asterism (Chevalier \& Ilovaisky 1991). Typical seeing for the nights
ranged between $1\farcs0$ and $1\farcs5$. 
The FOCAS image processing system (Valdes 1982) was used for
processing the images and to produce total magnitudes
and morphological classes. The magnitude limits were around
24.5 in $B$ and 23.5 in $R$.

\subsection{CCD drift scans at the Palomar 5-m telescope} 

Drift scans of the {\em Lockman} field 
through two filters were obtained on March 29,
1989 with the 4-Shooter camera (Gunn et al. 1987) at the cassegrain
focus of the 5-m Hale telescope. The detectors are four 800 x 800
Texas Instrument CCDs covering a net field of $8\farcm7$  on a side
with an image scale of $0\farcs335$ per pixel. The four CCDs image
a contiguous 2 x 2 matrix of sky and the chips are oriented such
that the readout directions of each detector are parallel. During
the observations, the instrument was rotated such that the CCD 
columns were oriented north-south. Each drift scan was started
by pointing the telescope at a position just south of the 
{\em Lockman} field, and then driving the telescope north accross 
the field at a rate of $0\farcm90$ per min. The leading CCDs recorded 
the sky through the 'wide $V$' $F555W$ filters employed in the HST Wide 
Field/Planetary Camera (cf. MacKenty et al. 1992), while near-infrared
filters ($F785LP$; MacKenty et al. 1992) were placed in front of the
trailing chips. The system response curves for these filters as well
as the photometric properties are given in Postman et al. (1996).
 
The CCDs were operated in Time-Delay-and-Integrate (TDI) mode at a rate
corresponding to the drift scan, resulting in an effective exposure 
of 293 s through each of the two filters. Except for 
the~$\approx$~$2\arcsec$ gaps in the center of each scan caused by the
4-Shooter beam-splitter, the entire field was 
covered with five drift scans 
whose centers were separated by 1.0 min in right ascension.
The seeing FWHM was approximately $2\farcs5$.
The initial processing of the drift scans followed the procedure
described in Schneider et al. (1994). The scans were
divided into a series of 816x800 images, with an 80 pixel overlap 
in the scan direction so every object of interest would be entirely
contained in one frame. A bias level, calculated from the extended
registers stored in the data, was subtracted from each frame, and a 
one-dimensional flat field was calculated for each detector by
median filtering the data. Temporal variations in sky brightness,
which appear as a change in background level in the scan direction,
were removed from each frame by fitting third-order polynomials
to the data.

Optical images were detected using the MIDAS package ``inventory''. 
The $3\sigma$ magnitude
limit was around 23.5 in $V$ and 22.5 in $I$.
The object positions were corrected for overall
systematic position errors along and accross the scan path using the
SKICAT scans of the POSS-II Schmidt plates (Weir et al. 1996).
We estimate the astrometric accuracy to be better than 0.5$\arcsec$.

\section{Optical spectroscopy}  

Given the accuracy of the PSPC positions, which were the basis for
the early phase of the identifications, and guided by the
simulations described in Paper I, we considered
any optical object within $15\arcsec$ from the position of the
X-ray source as a potential identification. We named optical
objects near the X-ray position A, B, \dots, usually in order of
increasing distance from the X-ray position.
In a few cases, the X-ray position moved in subsequent analyses
based on additional X-ray exposures,
and optical objects farther in the alphabet became involved. 
For each X-ray source, spectra were obtained of optical candidates,
more or less in order of brightness,
until a probable identification was made (cf. Sect. 4). Finding
charts and a full description of the spectra will be published
elsewhere (Lehmann et al., in prep.).

\subsection{Palomar spectra}  

Optical spectra for some of the brighter objects were obtained 
in February and December, 1992 with the 4-Shooter spectrograph 
(Gunn et al. 1987) at the 5-m Hale telescope. The spectrograph's
entrance aperture is a 1.5 x 100 $\arcsec$ slit and the detector an
800 x 800 Texas Instruments CCD. A 200 line mm$^{-1}$ transmission
grating produces spectra from 4500-9500\AA\ at a spectral resolution 
of 25\AA. The data were processed using the procedure given in
Schneider et al. (1994). 
Several of the approx. 20 spectra taken showed that the objects 
were active galactic nuclei (AGNs), which at the time provided assurance
that the celestial positions of the X-ray objects were approximately
correct. Spectra for most of these objects were subsequently obtained
with the Keck telescope (see below). For 10 objects (8B, 28B, 16A, 62A, 
20A, 27A, 9A, 25A, 41C and 52A), the Palomar spectra are the basis for 
their entry in Table 1 (cf. Sect. 4). These include the three galactic 
stars among the X-ray sources.

\begin{table*}
\caption[ ]{Photometric and spectroscopic properties of optical
identifications of the X-ray sources$^{\rm a}$}
\begin{flushleft}
\begin{tabular}{cccccccccccccc}
\noalign{\smallskip}
\hline
\noalign{\smallskip}
 $N_x$ & $S_x$ & name & $R$ &
$\alpha_{2000}$ & $\delta_{2000}$ & $\Delta$ pos & $\log f_x/f_v$ 
 & $G$ & $z$ & $M_V$ & $\log L_x$ & class & ID class \\
\noalign{\smallskip}
\hline
\noalign{\smallskip}
  28 & 19.90 &  28B & 18.2 & 10 54 21.3 & 57 25 44.3 &  1 H &  0.29 & g & 0.205 & -22.0 & 43.59 & AGN  & c      \\
   8 & 11.84 &   8B & 14.1 & 10 51 31.1 & 57 34 39.4 &  1 H & -1.58 & s &       &       &       & star & dMe    \\
   6 &  8.82 &   6A & 18.4 & 10 53 16.9 & 57 35 52.3 &  1 H &  0.01 & s & 1.204 & -25.6 & 44.89 & AGN  & a     \\ 
  32 &  6.50 &  32A & 18.1 & 10 52 39.6 & 57 24 31.7 &  1 H & -0.24 & s & 1.113 & -25.8 & 44.69 & AGN  & a      \\
  29 &  5.13 &  29A & 19.5 & 10 53 35.1 & 57 25 41.6 &  1 H &  0.22 & s & 0.784 & -23.6 & 44.25 & AGN  & b      \\
  31 &  3.57 &  31A & 20.0 & 10 53 31.8 & 57 24 53.8 &  0 H &  0.26 & s & 1.956 & -25.0 & 44.97 & AGN  & a      \\
  16 &  3.56 &  16A & 19.8 & 10 53 39.8 & 57 31 03.9 &  1 H &  0.18 & s & 0.586 & -22.7 & 43.81 & AGN  & c      \\
  56 &  3.52 &  56D & 18.9 & 10 50 20.2 & 57 14 21.7 &  1 R & -0.19 & s & 0.366 & -22.6 & 43.37 & AGN  & b      \\
  62 &  3.20 &  62A & 11.0 & 10 52 01.3 & 57 10 45.9 &  2 R & -3.39 & s &       &       &       & star & FG     \\
  37 &  2.54 &  37A & 19.6 & 10 52 48.2 & 57 21 17.4 &  2 H & -0.05 & s & 0.467 & -22.4 & 43.46 & AGN  & b      \\
  20 &  2.43 &  20C & 15.5 & 10 54 10.4 & 57 30 37.9 &  2 H & -1.71 & s &       &       &       & star & dMe    \\
   9 &  1.88 &   9A & 20.0 & 10 51 54.5 & 57 34 37.7 &  0 H & -0.02 & g & 0.877 & -23.4 & 43.92 & AGN  & b      \\
  41 &  1.87 &  41C & 17.9 & 10 53 18.7 & 57 20 43.9 &  5 H & -0.86 & g & 0.340 & -23.4 & 43.03 & Grp  & e      \\
  25 &  1.84 &  25A & 20.6 & 10 53 45.0 & 57 28 40.2 &  0 H &  0.21 & s & 1.816 & -24.3 & 44.61 & AGN  & a      \\
  42 &  1.69 &  42Y & 20.7 & 10 50 16.1 & 57 19 53.8 &  7 P &  0.22 & s & 1.144 & -23.2 & 44.13 & AGN  & a      \\
  48 &  1.64 &  48B & 19.9 & 10 50 46.2 & 57 17 33.1 &  1 R & -0.12 & g & 0.498 & -22.3 & 43.33 & AGN: & e      \\
  12 &  1.57 &  12A & 22.9 & 10 51 48.8 & 57 32 48.4 &  1 H &  1.06 & s & 0.990 & -20.7 & 43.96 & AGN  & d      \\
  59 &  1.50 &  59A & 16.9 & 10 53 24.8 & 57 12 30.7 &  5 P & -1.36 & g & 0.080 & -21.3 & 41.63 & AGN  & c      \\
  35 &  1.48 &  35A & 18.9 & 10 50 39.6 & 57 23 36.3 &  1 R & -0.56 & s & 1.439 & -25.5 & 44.29 & AGN  & a      \\
 117 &  1.46 & 117Q & 22.8 & 10 53 48.8 & 57 30 33.9 &  0 H &  0.99 & s & 0.780 & -20.3 & 43.70 & AGN  & d      \\
  27 &  1.39 &  27A & 20.3 & 10 53 50.3 & 57 27 09.2 &  0 H & -0.03 & g & 1.720 & -24.5 & 44.44 & AGN  & e      \\
  73 &  1.39 &  73C & 20.6 & 10 50 09.6 & 57 31 43.5 & 14 P &  0.09 & s:& 1.561 & -24.0 & 44.34 & AGN  & a      \\
  52 &  1.32 &  52A & 20.4 & 10 52 43.3 & 57 15 44.6 &  2 H & -0.01 & s & 2.144 & -24.8 & 44.63 & AGN  & a      \\
  11 &  1.27 &  11A & 23.0:& 10 51 08.4 & 57 33 45.4 &  0 H &  1.01 & g & 1.540 & -21.5 & 44.29 & AGN  & a      \\
  67 &  1.24 &  67B & 20.5:& 10 50 56.2 & 57 06 47.9 &  8 P & -0.48 & s & 0.550 & -21.9 & 43.29 & Grp  & e      \\
  26 &  1.20 &  26A & 18.7 & 10 50 19.8 & 57 28 12.2 &  6 P & -0.73 & g & 0.616 & -23.9 & 43.39 & AGN  & d      \\
  55 &  1.17 &  55C & 21.4:& 10 50 09.4 & 57 14 43.3 &  1 P &  0.34 & s:& 1.643 & -23.3 & 44.32 & AGN  & a      \\
   2 &  1.16 &   2A & 20.1 & 10 52 30.1 & 57 39 13.4 &  2 H & -0.19 & s & 1.437 & -24.3 & 44.18 & AGN  & a      \\
  54 &  1.12 &  54A & 20.3 & 10 53 07.4 & 57 15 04.6 &  2 H & -0.12 & s & 2.416 & -25.2 & 44.67 & AGN  & a      \\
  45 &  1.04 &  45Z & 21.1 & 10 53 18.9 & 57 18 50.0 &  2 H &  0.17 & s & 0.711 & -21.8 & 43.46 & AGN  & d      \\
  19 &  0.99 &  19B & 21.8 & 10 51 37.5 & 57 30 43.2 &  1 H &  0.42 & s & 0.894 & -21.6 & 43.66 & AGN  & b      \\
 504 &  0.96 &  51D & 20.2 & 10 51 14.5 & 57 16 15.5 &  1 R & -0.23 & s & 0.528 & -22.1 & 43.15 & AGN  & d      \\
  43 &  0.94 &  43A & 23.0 & 10 51 05.1 & 57 19 23.2 &  2 R &  0.88 & s & 1.750 & -21.8 & 44.28 & AGN  & a      \\
  36 &  0.92 &   &  &  & see text &  &  &  &  &  &  & -- & e    \\
  46 &  0.89 &  46A & 22.6 & 10 51 20.1 & 57 18 47.9 &  2 R &  0.70 & s & 1.640 & -22.1 & 44.20 & AGN  & a      \\
  61 &  0.79 &  61B & 20.8 & 10 51 26.3 & 57 11 31.1 &  7 P & -0.07 & g & 0.592 & -21.7 & 43.17 & AGN  & b      \\
  38 &  0.78 &  38A & 21.3 & 10 53 29.5 & 57 21 03.9 &  2 H &  0.12 & s & 1.145 & -22.6 & 43.79 & AGN  & a      \\
  60 &  0.78 &  60B & 21.6 & 10 52 48.5 & 57 12 06.0 &  3 P &  0.24 & g & 1.875 & -23.4 & 44.27 & AGN  & a      \\
  14 &  0.72 &   &  &  & see text &  &  &  &  &  & & -- & e     \\
  47 &  0.71 &  47A & 21.9 & 10 52 45.0 & 57 17 33.4 &  5 H &  0.32 & g & 1.058 & -21.9 & 43.68 & AGN  & a      \\
  30 &  0.70 &  30A & 21.5 & 10 52 57.3 & 57 25 07.1 &  1 H &  0.15 & s & 1.527 & -23.0 & 44.02 & AGN  & a      \\
  51 &  0.66 &  51L & 21.1 & 10 51 17.0 & 57 15 51.4 &  5 P & -0.03 & s & 0.620 & -21.5 & 43.14 & AGN  & d      \\
  17 &  0.62 &  17A & 20.3 & 10 51 04.0 & 57 30 54.0 &  2 H & -0.38 & s & 2.742 & -25.4 & 44.54 & AGN  & a      \\
 814 &  0.61 &  37G & 20.5 & 10 52 44.8 & 57 21 23.2 &  2 H & -0.31 & s & 2.832 & -25.3 & 44.57 & AGN  & a      \\
  84 &  0.60 &  &  &  & see text &  &  &  & &  &  & -- & e      \\
  77 &  0.59 &  77A & 21.7 & 10 52 59.3 & 57 30 30.2 &  1 H &  0.16 & s & 1.676 & -23.0 & 44.04 & AGN  & a      \\
  53 &  0.58 &  53A & 18.4 & 10 52 06.3 & 57 15 24.7 &  5 P & -1.17 & g & 0.245 & -22.2 & 42.22 & GAL  & e      \\
 116 &  0.57 & & & & see text & & & & & & & -- & e      \\
  23 &  0.56 &  23A & 21.9 & 10 52 24.7 & 57 30 09.6 &  1 H &  0.22 & g & 1.009 & -21.8 & 43.53 & AGN  & b      \\
  58 &  0.56 &  58B & 21.4 & 10 52 38.8 & 57 12 59.7 & 10 P &  0.02 & g & 0.629 & -21.3 & 43.08 & Grp   & e      \\
\noalign{\smallskip}
\hline
\end{tabular}
\end{flushleft}
\begin{list}{}{}
\item[$^{\rm a}$] For a description of the entries, cf. Sect. 3, and
for a discussion of the ID classes, cf. Sect. 4. 
\end{list}
\end{table*}

\subsection{Keck spectra}

Spectra for all remaining optical objects were taken with the Low
Resolution Imaging Spectrometer (LRIS) (Oke et al. 1995) at the 
cassegrain focus of the Keck 10-m telescope in February and
December 1995, April 1996 and April 1997. 
The slit width is either 1$\farcs$0 or
0$\farcs$7 and the detector is a back-illuminated 2048 x 2048
Tektronics CCD. A 300 line mm$^{-1}$ reflection grating produces
spectra from 3800-8200\AA\ at a spectral resolution of around 10\AA.
In order to minimize slit losses due to atmospheric selective
refraction, we generally rotated the instrument so that the projection 
of the slit on the sky was vertical.

The raw data frames were bias-subtracted and flatfield corrected.
The extraction of one-dimensional spectra from the two-dimensional
sky-subtracted spectra was done using the optimal extraction algorithm
for CCD spectroscopy described by Horne (1986). Third order polynomials 
were fitted to the lines of the He-Ar or Hg-Kr spectrum
observed before or after the object spectrum to determine the
wavelength scale. The flux calibration of the spectra was obtained
using secondary standard stars for spectrophotometry (Oke \& Gunn 1983).
An atmospheric correction function for the broad molecular
absorption bands (H$_2$O, O$_2$) was derived from the spectra of the
standard stars and applied to all other spectra.

\section{Optical identifications and source classifications}

In this section, we define systematically the process
followed in obtaining optical identifications of the
X-ray sources. In order to facilitate this description, we
introduce five identification classes (ID class), which categorize
the process of identification and classification used in each class. 

The primary goal is to obtain the optical identification of the
X-ray source, i.e., the correct association of the X-ray source with 
an optical object. Once the identification has been made, the redshift 
of the X-ray source is obtained from that of the optical object.
The classification of the source (star, galaxy, cluster of galaxies,
AGN, etc.)  is usually made on the basis of its optical properties.
For some ID classes, we will also use the X-ray properties in 
classifying the source. 

Since the surface density of optical objects that can be observed
is much larger than that of the X-ray sources in our survey, the 
identification depends on finding an optical object within the
X-ray error circle that has some unusual properties. In practice,
we find for about half of the sources an optical object with broad
emission lines, that we recognize as an AGN. We discuss these objects
below under ID classes $a-c$, and argue that AGNs are so rare
that their detection near the X-ray source 
constitutes an almost certain identification of the X-ray source.

Among the remaining sources, we find optical objects with [Ne\,{\sc v}]
or [Ne\,{\sc iii}] emission (ID class $d$). We argue that [Ne\,{\sc v}]
is indicative of soft X-ray emssion, confirming the identification
of the optical object with the X-ray source. For each of the two 
emission lines, we argue that they
are rarely observed and confirm the optical identifications.
On the basis of the Ne emission and the high X-ray 
luminosities, we classify the objects as AGNs.

For the remaining sources, we cannot find within the X-ray error
circle optical objects with unusual properties (ID class $e$).
For these sources we use the ratio of X-ray flux to optical flux 
$f_x/f_v$ as additional tool in our classification procedure.
This ratio is independent of distance and allows
distinguishing different classes of X-ray sources
(cf. Stocke et al. 1991). Once brighter optical objects 
(with smaller values of $f_x/f_v$) have been
eliminated as candidate identifications, we find for most of the
objects in ID class $e$ that the ratio $f_x/f_v$ is so large that 
the X-ray source cannot be a galaxy. 

Clusters and groups of galaxies show approximately the same range
of $\log f_x/f_v $ as AGNs (cf. Stocke et al. 1991). 
We do not see a rich cluster of galaxies 
on the CCD images at the position of 
most of the X-ray sources in ID class $e$. Therefore, among the 
known object classes typically associated with X-ray sources,
they could in principle be AGNs or groups of galaxies. The 
intragroup medium in poor groups of galaxies has been identified
as an important class of X-ray emitters through studies 
with ROSAT (Mulchaey et al., 1996). In these objects a small 
number of galaxies is surrounded by diffuse X-ray halos
with a typical linear diameter of 50-400 kpc and X-ray 
luminosities in the range $10^{42.5-43.5}$ erg s$^{-1}$.
Often the peak of the X-ray emission is not centered 
on any one galaxy, but in some cases a cooling flow is 
centered on a dominant galaxy (see also Stocke et al.,
1991). The local volume 
density of these groups (Henry et al., 1995) is large 
enough, that some higher redshift objects of this class 
are expected to show up in our survey. 
Similar objects at a redshift of 0.5 would have X-ray angular 
diameters (FWHM) of $10-50\arcsec$, and fluxes in the range 
$0.2-2 \times 10^{-14}$ erg cm$^{-2}$ s$^{-1}$. The optical 
galaxies in these groups would often be seen at $\sim$10\arcsec\ 
from the centroid of the X-ray emission. 

Table 1 provides a complete list of the X-ray sources of the
survey and the properties of their optical identifications.
The first two columns give the name of the X-ray source and its
flux in units of $10^{-14}$ erg cm$^{-2}$ s$^{-1}$ in the $0.5-2$
keV energy band (cf. Paper I). The next four columns give the name of 
the optical object identified with the X-ray source, its magnitude $R$, 
and its right ascension and declination at epoch 2000. 
The next column gives the distance of the optical object to the
X-ray source in arcsec and indicates whether the X-ray position is mainly
based on the 207 ksec PSPC pointing (P), the 1112 ksec HRI pointing (H)
or the 205 ksec HRI raster scan (R). Next we list the
ratio $f_x/f_v$ as defined by Stocke et al. (1991),
a morphological parameter $G$ that is either s = star or g = galaxy,
and the redshift $z$. The next two columns give the optical
absolute magnitude $M_V$ (assuming $V - R = + 0.22$, corresponding to a 
power law spectral index of $-0.5$) and the X-ray luminosity $L_x$ in units 
of erg s$^{-1}$ in the $0.5-2.0$ keV energy band (assuming an energy
spectral index of $-1.0$). 
In deriving these luminosities, we used
$H_o =$ 50 km s$^{-1}$ Mpc$^{-1}$ and $q_o =$ 0.5. 
The last two
columns give the classification of the object and the ID class. 
The ID class characterizes the identification and classification
procedure discussed in the next section.

Among the 50 X-ray sources in Table 1, we identify three as
galactic stars. For the
remaining objects the identification and classification
procedure is discussed in some detail below, according to its
ID class, which for each object is given in Table 1.
We show in Fig. 1 Keck LRIS spectra
illustrating the various ID classes discussed below.

\subsection{ID class $a-c$}
The ID classes $a-c$ are all based on the detection of broad emission
lines that are characteristic of AGNs. The ID class $a$ applies to
objects whose spectra show broad Mg\,{\sc ii} and C\,{\sc iii}]  
emission and, at sufficiently large redshifts, C\,{\sc iv} and  
Ly$\alpha$ emission. Most of the 21 objects in this class have high 
optical luminosities and are classical quasars (QSOs). All of the 21
are classified as active galactic nuclei (AGNs). 

ID class $b$ is assigned to 7 objects that show broad Mg\,{\sc ii} 
emission lines and ID class $c$ to 3 objects that exhibit broad emission 
from H$\beta$ (16A) or H$\alpha$ (28B, 59A). Among the class $b$ objects,
three (56D, 37A, and 9A) also show broad H$\beta$ emission. 

\begin{figure}
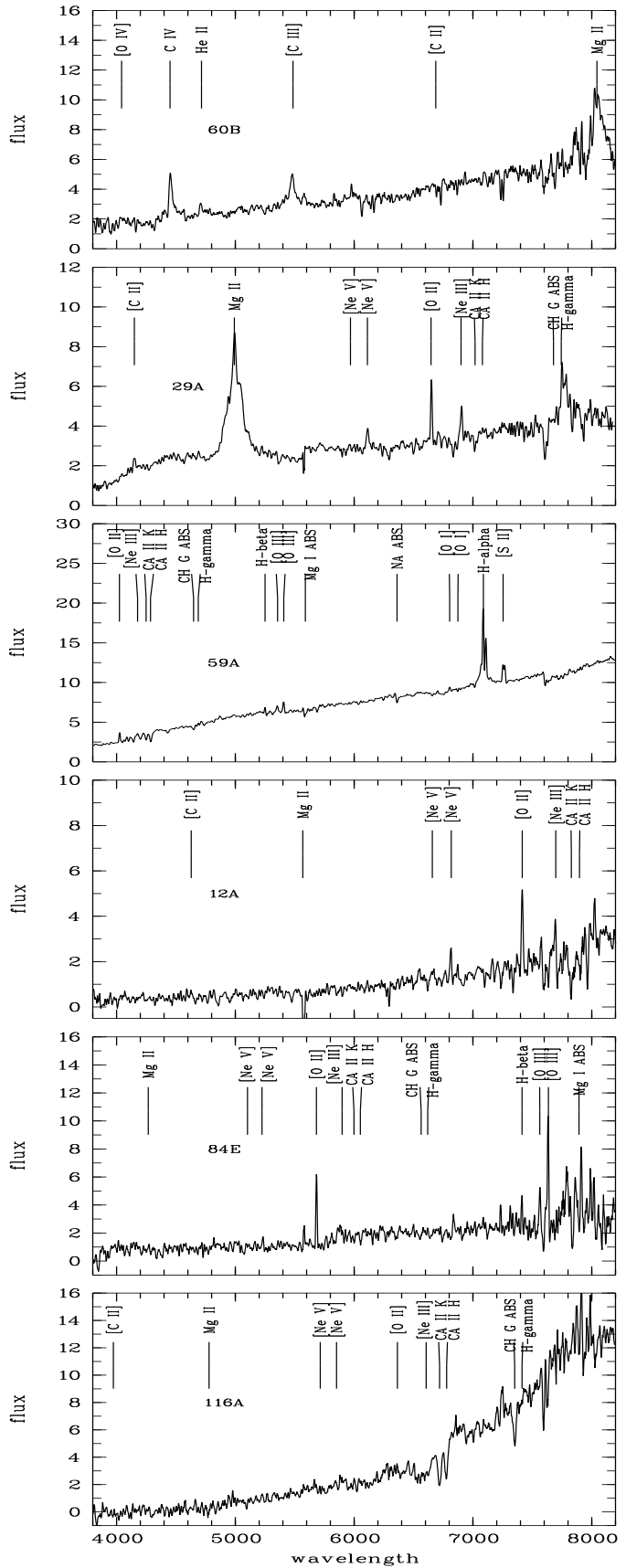

\begin{minipage}{90mm}
\psfig{figure=h0500.f1,bbllx=75pt,bblly=163pt,bburx=550pt,bbury=366pt,width=90
mm,height=37mm,clip=
}
\end{minipage}
\begin{minipage}{90mm}
\psfig{figure=h0500.f2,bbllx=75pt,bblly=163pt,bburx=550pt,bbury=366pt,width=90
mm,height=37mm,clip=
}
\end{minipage}
\begin{minipage}{90mm}
\psfig{figure=h0500.f3,bbllx=75pt,bblly=163pt,bburx=550pt,bbury=366pt,width=90
mm,height=37mm,clip=
}
\end{minipage}
\begin{minipage}{90mm}
\psfig{figure=h0500.f4,bbllx=75pt,bblly=163pt,bburx=550pt,bbury=366pt,width=90
mm,height=37mm,clip=
}
\end{minipage}
\begin{minipage}{90mm}
\psfig{figure=h0500.f5,bbllx=75pt,bblly=163pt,bburx=550pt,bbury=366pt,width=90
mm,height=37mm,clip=
}
\end{minipage}
\begin{minipage}{90mm}
\psfig{figure=h0500.f6,bbllx=75pt,bblly=130pt,bburx=550pt,bbury=366pt,width=90
mm,height=42mm,clip=
}
\end{minipage}
\caption[ ]{From top to bottom, Keck LRIS spectra of objects 
of ID class $a-e$, and an absorption-line galaxy at $z=0.71$}
\end{figure}

In all these cases, we believe that the object is an AGN. And
since AGN are both rare and generally strong X-ray sources, we
consider the optical object to be the identification of the 
X-ray source. This procedure may of course produce field AGNs that 
are not responsible for the X-ray emission. We estimate the 
number of such misidentifications as follows.
The surface density of broad-line AGNs with $B < \sim 22$
is around 115 deg$^{-2}$ (Zitelli et al. 1992).
This includes both quasars ($M_B < -23.0$) and Seyfert galaxies
($M_B > -23.0$). Extrapolating the observed counts to fainter 
magnitudes with a slope similar to that observed in the range
$20 < B < 22$, a total surface density of $ \sim 365$ AGNs deg$^{-2}$
has been estimated at $B < 23.5$ (Zamorani 1995).
For a $-0.5$ power law, $B = 23.5$ corresponds to $R \sim 23.0$, which is 
approximately the faint limit of the identifications listed in Table 1.
For a surface density of 365 deg$^{-2}$, the number of field AGNs 
expected in 50 circles of radius $15\arcsec$ is $ \sim 1$.   

\subsection{ID class $d$}

We use ID class $d$ to cover objects 
that do not exhibit the emission features shown in ID classes $a-c$,
but that show Ne emission lines. 
Four objects (12A, 117Q, 51D, and 51L) exhibit
[Ne\,{\sc v}] $\lambda$3426 emission. Since the ionization potential of
Ne$^{+++}$ is 97 eV, corresponding to soft X-ray emission,
the presence of [Ne\,{\sc v}] emission essentially confirms 
the optical identification of the X-ray source. (Among the
objects of ID class $a-c$ discussed above, we see [Ne\,{\sc v}]  
emission in the spectra of four of them.) 
With the identification of the optical object confirmed 
and therefore the redshift of the X-ray
source established, we find that the X-ray luminosity of all
four sources is above $10^{43}$ erg s$^{-1}$, far higher than that
of galaxies, and we conclude that they are AGNs.

Two objects (26A and 45Z) show relatively strong
[Ne\,{\sc iii}] $\lambda$3869 emission.
This feature, although seen also in field galaxies with emission
lines (cf. Hammer et al. 1997), is typically stronger 
in the spectra of quasars and Seyfert galaxies.
Among the 10 AGNs of ID class $a-c$ in which the [Ne\,{\sc iii}]
emission is in the accessible spectral range, we see it in 7 cases.
In all cases where we see [Ne\,{\sc v}]
emission, we also observe [Ne\,{\sc iii}] if it is accessible.
Object 26A also exhibits Fe\,{\sc ii} $\lambda$2600 absorption,
which has been seen in spectra of broad absorption line quasars
(Becker et al. 1997).
We accept the two objects as the optical identifications, and 
then (as for the [Ne\,{\sc v}] cases above) from the redshifts 
and the X-ray luminosities conclude that the objects are AGNs.

\subsection{ID class $e$}

In ID class $e$, we discuss the remaining objects that have 
galaxy counterparts without broad emission lines and 
without rare spectral features such as Ne emission lines. 
We have 10 objects with ID class $e$ in Table 1, of which four  
exhibit [O\,{\sc ii}] emission. Claims have been made (Griffiths 
et al. 1996, McHardy et al. 1997) that a significant fraction of
the weaker X-ray sources are narrow-emission line galaxies (NELG),
characteristically showing strong [O\,{\sc ii}] $\lambda$3727 emission,
and that these constitute a new population of X-ray sources.
Recent galaxy surveys have revealed 
that a large fraction of faint field galaxies show 
[O\,{\sc ii}] emission (cf. Hammer et al. 1997). Our own spectral work at 
Palomar and Keck has recorded spectra of 24 galaxies near the objects 
listed in Table 1 that have redshifts different from the optical
identifications. Among these field 
galaxies, 19 exhibit [O\,{\sc ii}] emission, i.e., 79\%
of field galaxies are NELGs. The detection of a NELG can therefore 
not be taken as evidence for the correct identification of an X-ray source.

Since the optical spectra of possible X-ray counterparts 
in ID class $e$ are undistinguished, we introduce two additional
criteria by which to identify and classify the objects. 
The first criterion is the ratio of X-ray and optical flux $f_x/f_v$ 
which is independent of distance (cf. Stocke et al. 1991).
In Fig. 2, we plot $\log f_x/f_v$ vs. the flux $S(0.5-2\,\rm{keV})$
for the 37 AGNs of ID class $a-d$ that 
we have identified so far, and for all AGNs and galaxies
in the Extended Medium-Sensitivity Survey (EMSS, Stocke et al. 1991).
Stocke et al. noted that essentially all AGNs
in the EMSS are in the range $-1.0 < \log f_x/f_v < 1.2$.
Among the 37 Lockman AGNs plotted, 36 have $\log f_x/f_v > -1.0$.
There is no evidence from Fig. 2 for evolution of the $f_x/f_v$
ratio of AGNs; for galaxies no or little effect of evolution
would be expected given their relatively small redshifts.

\begin{figure}
\begin{minipage}{90mm}
\psfig{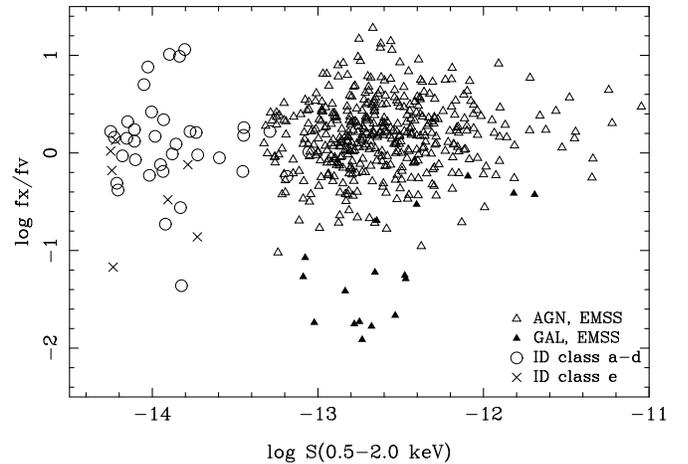}
\end{minipage}
\caption[ ]{Ratio of X-ray over optical flux plotted 
versus X-ray flux for Lockman and EMSS sources}
\end{figure}

The second criterion we considered is the angular extent of the X-ray 
emission. A rigorous determination of the X-ray extent for all sources
has not been part of our analysis scheme because it breaks down in the 
presence of significant confusion (see Paper I). Therefore the extent 
has been estimated {\it a posteriori} for the objects in question by fitting
a two-dimensional Gaussian to the HRI and PSPC images and comparing this 
to other nearby X-ray sources. Sizes are given in FWHM, approximately 
corrected 
for the size of the PSF. If a faint object is extended, it may be missed
by the HRI observations, which are less sensitive for low surface 
brightness emission. On the other hand, the HRI may pick up some
sharper structure within diffuse emission. Because of its softer response
it is actually more sensitive to detect cooling flows. As noted before,
optical galaxies in X-ray emitting poor groups would often 
be seen at positions outside the X-ray error circle. For an extended X-ray
source we consider the detection of two or more galaxies at the same
redshift as supporting the group identification.

In summary, while the X-ray emission from AGNs and single galaxies 
should be point-like and centered on the optical counterpart, clusters 
and groups of galaxies may have extended X-ray emission (and therefore
reduced HRI fluxes relative to PSPC fluxes), that is displaced
from one or more galaxies. The discussion given below for each of the 
sources takes into account these additional criteria in order to 
illuminate whether it is more likely to be an AGN, a group of galaxies 
or something else.

Object 41 is extended by about 22$\arcsec$ in both directions on the
1112 ksec HRI exposure and by 50$\arcsec$ in the PSPC exposure. 
Optically, a dense group of galaxies is visible, of which 41C 
is the brightest member. Several other galaxies are at the same 
redshift as 41C. We classify the object as a
group of galaxies, although it could be a cluster.

Object 48 is not extended in the PSPC exposure. It is seen as a point source 
in the HRI raster scan, well centered on the galaxy 48B but is not covered by 
the deep HRI pointing. Two other galaxies, at 9$\arcsec$ 
and 85$\arcsec$ from the X-ray source, have the same redshift as 48B. 
The object has a high $\log f_x/f_v $ ratio. Considering the
excellent coincidence of the optical and the HRI position and the
absence of evidence for an extent in X-rays, we consider 48B to be
the optical identification. On the basis of its X-ray luminosity and
$f_x/f_v$ ratio, we classify this object as an AGN, although this
classification has to be considered uncertain.

Object 27 is a complex case. The PSPC X-ray position is consistent with
that of the optical quasar 27A. Early parts of the 1112 ksec HRI exposure
show a source 10$\arcsec$ SE of 27A but the
more recent HRI exposures produce a source position less than 1$\arcsec$
from 27A. Apparently, there are two variable sources involved. Since
the survey is defined by the epoch of the PSPC observations, we
identify the X-ray source with 27A.

Object 67 appears complex in the PSPC image. It is near the edge of
the field, at 17$\farcm$7 from the center. It is barely detected on
the HRI raster scan, suggesting a substantial extension in X-rays. 
The galaxy 67B has $\log f_x/f_v $=-0.48. 
Three other galaxies in the error box have the same redshift as 67B. 
We classify the object as a group.

Object 36 is another case where variability plays a role. The source 
36 is based on the PSPC exposures. The deep HRI survey shows an X-ray 
source (designated as number 870) about 15$\arcsec$ to the south of 36. 
Near this position,  we found the optical object 36F, which is an AGN at 
redshift 0.8, and clearly the identification. However, the HRI raster 
scan, which only recently produced accurate positions after an 
astrometrical solution provided by the ultradeep HRI pointing 
(see Paper I), clearly confirms the PSPC position and does not show 
an object near 36F. It thus appears that there are two X-ray sources 
and that 36F is strongly variable. Since, as stated, our survey is 
defined by the epoch of the PSPC observations, source 870 is not part 
of the current survey. We have resumed work to identify source 36, but 
it is incomplete so far.

\begin{table*}
\caption[ ]{Comparison of ROSAT PSPC surveys}
\begin{flushleft}
\begin{tabular}{ccccccccccc}
\noalign{\smallskip}
\hline
\noalign{\smallskip}
Survey$^{\rm a}$ & Area$^{\rm b}$ & T$^{\rm c}$ 
& $S_{lim}^{\rm d}$ & $N^{\rm e}$ & AGN & GAL 
& unid. & FC$^{\rm f}$ & FL$^{\rm g}$ & FU$^{\rm h}$  \\
\noalign{\smallskip}
\hline
\noalign{\smallskip}
 CRSS & 3.9 & $>$6    & 2.0 & 123 & 55\%  &10\%&  &  &  &   \\
 DRS  & 1.4 & $21-49$ & 0.3 & 194 & 55\%  &10\%&18\%&8\%&38\%&13\% \\
 NEP  & 0.2 &  79     & 1.0 &  20 & 65\%  & 0\%&10\%& 5\%& 4\%& 5\% \\
 UKDS & 0.2 & 110   & 0.2 &  70 & 46\%  & 26\% &16\%&8\%&41\%&26\% \\
 RDS  & 0.3 & 207  & 0.5 &  50 & 78\% &  2\%& 8\%& 6\%&8\%& 5\% \\
\noalign{\smallskip}
\hline
\end{tabular}
\end{flushleft}
\begin{list}{}{}
\item[$^{\rm a}$] cf. text for explanation of acronyms 
\item[$^{\rm b}$] approximate, in square degrees  
\item[$^{\rm c}$] PSPC exposure time, in ksec
\item[$^{\rm d}$] limiting flux $S(0.5-2\rm{keV})$ in units
 of $10^{-14}$ erg cm$^{-2}$ s$^{-1}$
\item[$^{\rm e}$] number of X-ray sources detected  
\item[$^{\rm f}$] fraction of output sources that are contaminated,
 based on simulations  
\item[$^{\rm g}$] fraction of input sources that are lost,
 based on simulations  
\item[$^{\rm h}$] fraction of output sources that are unidentifiable,
 based on simulations  
\end{list}
\end{table*}

Object 14 appears as a point source in both the PSPC and the 1112 ksec
deep HRI exposures. The optical object nearest to the HRI position is
a galaxy at a distance of 7$\arcsec$. It has $R=22.5$, $z=0.546$, and 
its spectrum has no broad emission lines or Ne emission. 
The interpretation of this object depends critically on whether it is 
strictly a point source or has an X-ray extension of, say, 20$\arcsec$
that might well have escaped detection. If the source is extended,
the galaxy might be a member of a group of galaxies that is the X-ray
source. If on the other hand, the X-ray source is strictly a point
source, then the galaxy is unlikely to be the identification: simulations 
discussed in Paper I show that the probability that an HRI point 
source of this X-ray flux would appear at a distance of 7$\arcsec$ 
in our survey is around 2\%. Since a Keck CCD-image shows that there is 
no object 
brighter than $R=25$ in the HRI error circle, the single X-ray source 
would have $\log f_x/f_v > 1.6$, which would be truly outstanding.
We cannot decide between a group of galaxies and an X-ray intense
point source at the present time.
 
Object 84 appears to be a case similar to that of object 14.
The PSPC and HRI images are probably not extended. Two galaxies at 
about 14$\arcsec$ from the X-ray source have the same redshift.
The spectrum of the brighter galaxy (84E, $R=21.6$, $z=0.525$)
is illustrated in Fig. 1. 
There is no object in a deep $R$-band Keck image within the HRI error 
circle. The case for a group of galaxies is somewhat stronger here,
since we do have two galaxies with the same redshift. On the other hand, 
if object 14 is
a single object with large $f_x/f_v$ then 84 might be another such case.
We cannot decide between a group of galaxies and an X-ray intense
point source at the present time.

Object 53A shows H$\alpha$ emission with FWHM = 321 km/sec consistent
with instrumental broadening and it has $\log f_x/f_v = -1.17$.
Accordingly, we classify the optical object as a galaxy. At the
redshift of the galaxy, the X-ray luminosity ($\log L_x = 42.2$) is
low for an AGN and rather high for a galaxy. The galaxy is located
about midway between two bright M-type stars (neither of which
is an important contributor to the X-ray emission). 

Object 116 is extended by about 50$\arcsec$ EW on the PSPC exposure.
It is marginally seen on the 1112 ksec HRI exposure and probably extended.
Three galaxies at distances of 3$\arcsec$, 8$\arcsec$ and 10$\arcsec$
from the center of the PSPC error box, however, all have 
different redshifts (0.708, 0.408 and 0.610, respectively).
We cannot present an optical identification at present.

Object 58 appears extended in the EW direction on the PSPC exposure,
by about 40$\arcsec$. The object is not detected on the 1112 ksec HRI
exposure, consistent with its PSPC extent. At a distance of 
14$\arcsec$ from the PSPC position there is the galaxy 58B which
has $\log f_x/f_v = -0.02$. Another galaxy at a distance of 11$\arcsec$ 
from the center of the error box is at the same 
redshift. We classify the object as a group of galaxies.

\section{Comparison with other ROSAT PSPC surveys}

Several medium and deep surveys conducted with the PSPC on board
ROSAT have been published in the past several years. We are
interested in comparing our results with those surveys in
which optical identification and spectroscopy have been carried
out for the majority of the detected X-ray sources. These are:
\begin{itemize}
  \item The Cambridge-Cambridge ROSAT Serendipity Survey
        (CRSS, Boyle et al. 1995)
  \item A deep ROSAT survey (DRS, Georgantopoulos et al. 1996)
  \item The ROSAT North Ecliptic Pole Deep Survey 
        (NEP, Bower et al. 1996)
  \item The UK ROSAT deep field survey (UKDS, McHardy et al. 1997)
\end{itemize}

We list in Table 2 the properties of the surveys (the present
survey is denoted as RDS), as well as the
fractions identified as AGNs and galaxies. In general, the area 
covered by a given survey varies with limiting flux. The limiting 
fluxes $S_{lim}$ given in Table 2 are those mentioned in the 
quoted references and generally refer to the weakest X-ray 
sources in the sample.

The fraction of sources identified as AGNs in the RDS is larger
than that in all other surveys. This 
is partly due to the high quality of the Keck LRIS spectra and
partly due to our classification procedure. In particular, we
have used evidence based on the presence of [Ne\,{\sc v}] or 
[Ne\,{\sc iii}] emission, together with the high X-ray luminosity to
classify 6 objects (12\%) as AGNs that otherwise might have been
classified as galaxies.

We can carry out a direct comparison of the RDS with the 
two shallower surveys. Based on the CRSS AGN and galaxy percentages, 
we would expect to find among the 11 RDS sources above 
$2\, 10^{-14}$ erg cm$^{-2}$ s$^{-1}$ six AGNs and one galaxy,  whereas 
we observe eight AGNs and no galaxies. Similarly, based on 
the NEP AGN percentage, we would expect in the RDS above 
$10^{-14}$ erg cm$^{-2}$ s$^{-1}$ 20 AGNs, whereas we observe 25.
We conclude that the percentages of AGN and galaxies for the CRSS, the
NEP and the RDS are not significantly different.

The DRS and the UKDS have lower quoted limiting fluxes than the
RDS, even though their exposure times are shorter. Given the
simulations discussed in Paper I, this raises concerns about the
effects of confusion. We show in Table 2 the three F-factors
derived from simulations relevant to the different surveys, as
described in Paper I. The contamination factor $FC$ 
represents the fraction of output sources that have a flux
in excess of 1.5 times the nearest input source 
(if within 15\arcsec) augmented by 3$\sigma$. In practice, more
than half the photons of these output sources did not originate
from the nearby input source. The loss factor $FL$ is the fraction
of input sources that do not have an output source within
$10-15$\arcsec. These input sources are not detected at all or
have lost their identity
and in practice cannot be optically identified. The factor $FU$
is the fraction of output sources that has no input source within
15\arcsec. It therefore represents the unidentifiable fraction
of the detected sources.

The loss factors $FL$ for the DRS and the UKDS show that
38 and 41\%, respectively, of the input sources (i.e., the 
real sources in the field) are lost. In contrast,
the NEP and the RDS lose only 4 and 8\% of the input sources.
Actually, since in the RDS the 1112 ksec HRI survey covers about 2/3
of the $R<12\farcm5$ PSPC field, the effective values of $FC$ and $FL$
are less than those given in Table 2.

Among the sources that are detected, the factor $FU$ for the UKDS 
suggests that 26\% should be unidentifiable. McHardy et al. (1997)
state that 11 out of 70 sources, or 16\%, are unidentified.
The discrepancy between these two percentages raises the question
whether some of the unidentifiable sources in the UKDS may have been 
identified incorrectly. We suspect that this is the case, and that
the misidentifications involve the NELGs which are claimed to be
a new population of X-ray sources (Griffiths et al. 1996, McHardy 
et al. 1997). As discussed in Sect. 4.3, NELGs make up such a large
fraction of faint field galaxies that the presence of a NELG in the
X-ray error circle cannot be taken to confirm its identification of the
X-ray source.

\section{Discussion and conclusion}

Our procedure to identify and classify the weak X-ray sources
in the RDS depends on the angular extent of the object.
For stars and AGNs (cf. ID classes $a-d$) 
which are point sources both in X-rays and
optically, we require positional agreement between the X-ray and
the optical source. As can be seen for the objects of ID class $a-d$ 
in Table 1, the average offset between X-ray and optical positions is
$6\arcsec$ for PSPC positions and less than $2\arcsec$ for HRI positions.  
Identifications are based, first on the presence of broad emission 
lines, second on the combination of the presence of [Ne\,{\sc v}] 
emission and high X-ray luminosity, and third on the presence 
of relatively strong [Ne\,{\sc iii}] emission.
We argue that the presence of one or more broad emission 
lines in the spectrum of an optical object near the X-ray source 
signifies an AGN, and that these are sufficiently rare that 
this essentially confirms the identication. The presence of
[Ne\,{\sc v}] is taken to be indicative of soft X-ray emission, 
confirming the optical object as the
identification of the X-ray source, and the high X-ray luminosity
derived from the redshift signifies an AGN.
Given the rare ocurrence of strong [Ne\,{\sc iii}] emission, 
its presence is also considered to confirm the identification.

For the remaining 10 objects, of ID class $e$, the identification 
procedure is quite different. Many of these objects have extended 
X-ray emission and some have a low HRI flux or a positional offset
between HRI and PSPC. X-ray variability is seen in two cases,
complicating the identification procedure. 
Generally, the ratio $f_x/f_v$ is fairly large, 
suggesting that the object is an AGN or a group or
cluster of galaxies. We take evidence for X-ray source extension 
and the presence of several galaxies
at the same redshift as suggestive of identification with a group
of galaxies, even though we have not systematically explored
galaxies near the X-ray source. 

Applying these classification rules, we find 39 AGNs, 3 groups of 
galaxies, 1 galaxy and 3 stars. Among the AGNs, 16 have
$M_V < -23.3$, corresponding to $M_B < -23.0$ which according to
the luminosity criterion of Schmidt \& Green (1983) would qualify them
as quasars, while the other 23 AGNs according to their optical
luminosities are Seyfert galaxies. Among the four sources that remain 
unidentified, the observations for object 36 are not complete,
while source 116 is most likely a group of galaxies. The most intriguing 
X-ray sources are objects 14 and 84, for which there is no evidence
for source extension and where the ratio $f_x/f_v > 40$, larger than
any shown in Fig.~2. However, if these sources have modest extension
that could have remained undetected so far, they could still be
groups of galaxies, as discussed in Sect. 4.3.

Roche et al. (1995), Shanks et al. (1996) and Almaini et al. (1997)
find evidence for a significant cross-correlation between $B<23$
galaxies and weak unidentified X-ray sources. Almaini et al. find
that $B<23$ galaxies account for 23\% of the X-ray background at
1 keV. This finding is qualitatively consistent with the substantial
fraction of AGNs which are morphologically classified as galaxies
in our survey (see Table 1). Further analysis is required to
understand the relationship between these results in detail.

The fraction of X-ray sources
with reliable optical identifications and redshifts in the RDS is 
higher than that in any previously published X-ray survey. This
is a consequence of the realistic X-ray flux limits we used based
on simulations, the role of deep HRI exposures for many of the
sources in the central part of the PSPC field and the high 
quality of most of the Keck spectra obtained with LRIS. 
While the survey is based on sources selected above given flux
limits based on PSPC exposures, the effect of the subsequent
HRI imaging was considerable. In the early
phase of the identification work, X-ray positions were based on
PSPC images. In most cases, the identifications obtained
were confirmed later by the much more accurate HRI positions.
In two cases, the HRI managed to resolve confusion in the PSPC
image, resulting in each case in an object pair, 37A+37G and 51D+51L,
respectively. The HRI exposures also played a significant role in
the realization that a number of X-ray sources are not associated with 
single objects but rather are possible groups of galaxies 
(e.g., object 58), either by directly confirming their extent 
or showing an offset between X-ray and optical positions. 

We discuss claims in the literature for the appearance of a new 
population of weak X-ray sources in the form of narrow-emission 
line galaxies (NELG). We note that these claims are made on the 
basis of surveys that according to our simulations are strongly 
affected by confusion, leading to a substantial number of 
detected sources that are unidentifiable. Given that a large 
percentage of faint field galaxies are actually NELGs, we suggest 
that some of the unidentifiable sources identified as NELG are
field galaxies that have no relation to the X-ray source.
We do not claim that there cannot exist a new population among the
weaker X-ray sources. Given the natural uncertainties that accompany
the identification of objects of ID class $e$, such a new population
could only be found in surveys with conservative flux limits in
which confusion plays a minor role.

We will address in a subsequent communication the luminosity
functions of the extragalactic X-ray sources and the origin of the
X-ray background at energies below 2 keV.

\begin{acknowledgements}
The ROSAT project is supported by the Bundesministerium
f\"ur Forschung und Technologie (BMFT), by the National
Aeronautics and Space Administration (NASA), and the Science
and Engineering Research Council (SERC). The W. M. Keck Observatory
is operated as a scientific partnership between the California
Institute of Technology, the University of California, and the
National Aeronautics and Space Administration. It was made possible
by the generous financial support of the W. M. Keck Foundation.
We thank the Institute for Astronomy of the University of Hawaii,
Palomar Observatory, and the National Optical Astronomy 
Observatories for grants of observing time. We thank Bev Oke and Judy
Cohen for their efforts in the construction of LRIS,
E. Wyckoff and B. McLean for assistance in the photometric
and astrometric reduction of the CCD images taken with the University
of Hawaii 2.2 m telescope, and R. de Carvalho and G. Djorgovski for 
providing us with data from the POSS-II catalogue in our field.
We thank an anonymous referee for carefully reading the manuscript 
and suggesting important improvements.
M.S. thanks the Alexander von Humboldt-Stiftung for a Humboldt Research
Award for Senior U.S. Scientists in 1990-91; and the directors of the
Max-Planck-Institut f\"ur extraterrestrische Physik in Garching and of
the Astrophysikalisches Institut Potsdam for their hospitality.
This work was supported
in part by NASA grants NAG5-1531 (M.S.), NAG8-794, NAG5-1649,
and NAGW-2508 (R.B. and R.G.), NAG5-1538 and NAG8-1133 (R.B.),
and NSF grants AST86-18257A02 (J.G.) and
AST95-09919 (D.S.). G.H. acknowledges the DARA grant 
FKZ 50 OR 9403 5. G.Z. acknowledges partial support by the Italian 
Space Agency (ASI) under contract ASI 95-RS-152.

\end{acknowledgements}

\end{document}